\documentclass[aps,pre,reprint]{revtex4-2}
\usepackage{graphicx,amssymb,amsmath}
\usepackage{color}
\usepackage[dvipsnames]{xcolor}
\begin{document}
\title{Electrophoretic mobility of a water-in-oil droplet separately affected \\by the net charge and surface charge density}
\author{Yuki Uematsu}
\email[]{uematsu.yuki@phys.kyushu-u.ac.jp}
\affiliation{Department of Physics, Kyushu University, Motooka 744, Fukuoka 819-0395, Japan}
\affiliation{PRESTO, Japan Science and Technology Agency, 4-1-8 Honcho, Kawaguchi, Saitama 332-0012, Japan}
\author{Hiroyuki Ohshima}
\affiliation{Faculty of Pharmaceutical Sciences, Tokyo University of Science, 2641 Yamazaki Noda, Chiba 278-8510, Japan}
\date{\today}

\begin{abstract}
Water-in-oil emulsions and droplets exhibit completely different physico-chemical properties compared to oil-in-water emulsions and droplets.
Thus, directly applying a standard theoretical model to water-in-oil systems cannot describe these anomalous properties. 
Here, the electrophoretic mobility of a water-in-oil droplet is analytically investigated using Debye-H\"uckel linearization and neglecting Marangoni effect. 
The resulting electrophoretic mobility is shown to be separately dependent on the net charge of the droplet and the surface charge density at the droplet interface. 
Furthermore, when the net charge is negligible, electrophoretic mobility is proportional to the surface charge density with a negative coefficient. This indicates that the internal electric double layer inversely contributes the electrophoresis. 
This theory is applied to experimental data of water-in-oil emulsions and droplets in the literature, and qualitative and quantitative verification of the theory is discussed. 
\end{abstract}

\maketitle

\section{Introduction}

Emulsion stability is essential for chemical processing and basic surface science \cite{Bancroft1912, Griffin1949,Griffin1954,Winsor1948,Davies1957,Binks1995}, and it can be partially explained by the Derjaguin-Landau-Veerwey-Overbeek (DLVO) theory \cite{Derjaguin1941,Verwey1948}.
However, emulsions can coarsen via coalescence \cite{Langevin2019} and Ostwald ripening \cite{Lifshitz1959,Wagner1961}, which are unique to liquid droplets \cite{Binks1995}, in addition to sedimentation, creaming, and flocculation \cite{IsraelachviliBook,Russel1989}.
Ionic and non-ionic surfactants are typically used to obtain stable emulsions, and the stability of emulsions are partially determined by the chemical nature and partition of the surfactants, as well as the interactions between the droplets \cite{Binks1995}.
Although the composition of the two immiscible liquids is one of the factors that control the continuous phase (water or oil), the hydrophile-lipophile-balance (HLB) of the stabilizing surfactant plays a crucial role in determining the continuous phase and phase coexistence \cite{Griffin1949,Griffin1954,Davies1957,Binks1995}.
Surfactant-free emulsions have been considered unstable after emulsification due to creaming (sedimentation), flocculation, coalescence, and Ostwald ripening \cite{Binks1995}.
However, recent experimental studies have shown that surfactant-free emulsions are stable for some time \cite{Marinova1996,Sakai2002,Beattie2004,Sakai2008,Beattie2009, Roger2012, Richmond2019, Roke2020, Richmond2020}, and the stability of such surfactant-free oil-in-water emulsions is believed to be determined by the electrophoretic mobility (i.e., zeta potential) of the droplet interface \cite{Richmond2019,Roke2020,Richmond2020}. 

To date, most of the studies mentioned in the literature have focused on oil-in-water emulsions, and studies on water-in-oil emulsions are less common due to their low stability.
However, compared to oil-in-water emulsions, water-in-oil emulsions and droplets exhibit unique physicochemical properties \cite{Hermanie1952, Overbeek1959-1,Overbeek1959-2,Rigole1965,Kent2001,Ushikubo2014,Sakai2018,Sakai2021}.
In water-in-oil emulsions,
the continuous phase has a low dielectric constant, and the application of the DLVO theory to the stability is not straightforward \cite{Overbeek1959-1,Overbeek1959-2}.
The sign and the magnitude of the electrophoretic mobility of surfactant-stabilized water-in-oil emulsions has been measured \cite{Hermanie1952,Overbeek1959-1, Rigole1965, Kent2001, Ushikubo2014}. However, the relationship between emulsion stability and the charge of the stabilizing surfactants remains unclear \cite{Overbeek1959-1}.
For example, water-in-benzene emulsions show positive electrophoretic mobility even when anionic stabilizers are used \cite{Overbeek1959-1}.
Furthermore, surfactant-free water-in-oil emulsions were recently obtained using ultrasound irradiation \cite{Sakai2018, Sakai2021}, and they are also stable for a while \cite{Sakai2018}.  
Nevertheless, to the best of our knowledge, the electrophoretic mobility of surfactant-free water-in-oil emulsions has not been reported. 

As in water-in-oil emulsions, the electrophoretic mobility of water-in-oil droplets has been directly measured by pipetting a water droplet into the oil and applying a strong DC electric field \cite{Kang2008,Im2011,Hamlin2012,Esmaeil2012,Mesquida2013,Choi2013,Mesquida2014, Im2017, Im2020}.
Immediately after pipetting, the water droplet somehow has a positive electrophoretic mobility \cite{Mesquida2013,Choi2013}.
In these studies, the net charge of the droplet $Q$ is obtained from electrophoretic mobility $\mu$ by equating $\mu = Q/\gamma$, where $\gamma$ is the hydrodynamic friction coefficient depending on the model (e.g., Hadamard-Rybczynski friction \cite{Hadamard1911,Rybczynski1911}, Stokes friction \cite{Landau1987}, and the wall-effect-included friction coefficient \cite{Happel1973,Im2020}).
From a theoretical viewpoint, the application of $\mu=Q/\gamma$ is valid for the point-charge limit as known in the H\"uckel formula for the electrophoretic mobility of a charged solid sphere \cite{Huckel1924}.
Further, the convection induced by the electric double layer inside the water droplet should significantly contribute the electrophoretic mobility of the water-in-oil droplet, and thus, the effective charge converted from the electrophoretic mobility, defined by $Q_\mathrm{eff}=\gamma \mu$, should be different from the net charge $Q$.
Many theories for calculating the electrophoretic mobility of liquid droplets have been constructed thus far \cite{Booth1951,Saville1991,Ohshima2003,Squires2011,Levich1945,Levich1962,Levine1973,Ohshima1984,Squires2011,Yariv2013,Hill2020,Hill2020_2,Hill2021}.
Linear approximation in both electrostatics and applied electric field was first used for the droplet electrophoretic mobility whereas the surface charge gradient (Marangoni effect) was neglected \cite{Booth1951}. 
With fully nonlinear treatment in electrostatics, the electrophoretic mobility of a droplet has been numerically obtained for more general cases including surface charge gradient \cite{Saville1991}.
However, experimental comparison has been limited \cite{Sengupta1968, Brooks1984} except for mercury droplets \cite{Levich1962} and surfactant-stabilized oil-in-water emulsions \cite{Hill2020,Hill2020_2,Hill2021}, and more specific theory and experimental comparison for water-in-oil droplets have not been done yet. 
Therefore, a simple formula for the electrophoretic mobility of a water-in-oil droplet is necessary to discuss experimental data in the literature.

In this study, we reformulate the theory on the electrophoresis of droplets using Debye-H\"uckel linearization with neglecting the surface charge gradient \cite{Booth1951} and calculate the electrophoretic mobility of a water-in-oil droplet as a function of net charge and surface charge density.
Because our results are based on linear electrostatics, the effective charge deduced from our formula is not quantitatively correct compared to the model including nonlinear electrostatics [46,54-56].
Therefore, adopting our formula to interpret experimental data needs great care.
Nevertheless, our results are still significant because our theory reveals the uniqueness of the electrophoretic mobility of water-in-oil droplets compared to oil-in-water droplets or solid particles.
For the conventional theory of electrophoresis of solid particles, the most important parameter is the surface charge density (or zeta potential) \cite{Henry1931,OBrien1978}.
However, the water-in-oil droplet has an additional parameter: the net charge of the droplet.
When the oil phase is insulating, the net charge of the droplet can be considered a parameter independent of the surface charge density.
Using the derived electrophoretic mobility formula, we discuss the experimental results for water-in-oil emulsions and a water droplet in insulating oil \cite{Overbeek1959-1, Kang2008, Im2011, Choi2013, Mesquida2013, Mesquida2014, Im2017, Im2020}.
According to our theory, the positive electrophoretic mobility of a water-in-benzene emulsion stabilized by an anionic surfactant is naturally explained by the negative linear coefficient of the electrophoretic mobility with respect to the surface charge density. 
Furthermore, our theory suggests that the effective charge extracted from the electrophoretic mobility approaches a half of the total net charge of the droplet in the limit of large radius, whereas the experiment indicates that they are identical. 
However, to date, experimental studies have been inconclusive due to the difficulties in independent and accurate measurements of the total net charge \cite{Ristenpart2017} and the hydrodynamic friction coefficient \cite{Im2020}.
We hope that this paper helps explain the unique electrokinetics of water-in-oil emulsions and droplets.

\section{Model}
The model is similar to the work of Booth \cite{Booth1951}.
The analytic calculation of the electrophoretic mobility for a water-in-oil droplet is performed with the following assumptions. 
(1) A liquid droplet with radius $R$ is fixed at the center of the spherical coordinate and surrounded by another liquid.
The interface is fixed when external flow and electric fields are applied.
(2) The electrostatic potential is the sum of the solution of the linearized equilibrium Poisson-Boltzmann equation and the potential of an insulating droplet under an external electric field.
This is equivalent to neglecting the relaxation effect in the case of spherical solid particles \cite{OBrien1978}. 
In the case of liquid droplets, the external electric field or flow induces the surface charge gradient, and thus, Marangoni stress contributes to the force balance.
Solving fully nonlinear electrostatics and including Marangoni effect were performed in earlier works \cite{Saville1991,Hill2020,Hill2020_2,Hill2021}, but in this paper nonlinear electrostatics and Marangoni effect are neglected for simplicity. 
(3) The inertia terms in the electrohydrodynamic equations are neglected.
This is equivalent to concerning the electrophoretic velocity linear with the applied electric field.
(4)  
The internal and external liquids are incompressible, and the dielectric constants and viscosities are represented by $\varepsilon_\mathrm{in}$, $\varepsilon_\mathrm{out}$, $\eta_\mathrm{in}$, and $\eta_\mathrm{out}$. 
Additionally, the bulk salt concentration of the external liquid is zero, and the ion partition across the interface is neglected.
(5)
We do not explicitly consider ionic and non-ionic surfactants (or equivalently impurities) at the interface.
These surface-active molecules significantly alter the surface tension and surface charge density, and thus, they affect the hydrodynamic and electrostatic boundary conditions.
Instead, the surface charge density is infinitesimal (Debye-H\"uckel linearization) and it remains uniform and constant under external flow and electric fields (neglecting Marangoni effect).
Additionally, slip boundary condition (continuity of the tangential velocity and stress at the interface) is used.
The validity of these assumptions is discussed further in appendix.

Under these assumptions, the steady-state low-Reynolds-number electrohydrodynamic equations are as follows:
\begin{eqnarray}
\varepsilon_0 \nabla^2\psi &=& \left\{\begin{array}{ll}
-\rho(\boldsymbol{r})/\varepsilon_\mathrm{in}  & \textrm{for }r<R\\ 
-\rho(\boldsymbol{r})/\varepsilon_\mathrm{out}  & \textrm{for }r>R\\ 
\end{array}\right.,\\
\nabla\cdot \overleftrightarrow{\boldsymbol{\sigma}}&=&0, \label{eq:2}\\
\nabla\cdot\boldsymbol{u} &=& 0,\label{eq:3}\\
\nabla \cdot \boldsymbol{j}_+  &= & 0, \label{eq:4}\\
\nabla \cdot \boldsymbol{j}_-  &= & 0, \label{eq:5}
\end{eqnarray}
where $\psi$ is the local electrostatic potential, $\rho(\boldsymbol{r})$ is the ionic charge density, $\overleftrightarrow{\boldsymbol{\sigma}}$ is the stress tensor, $\boldsymbol{u}$ is the velocity field, and $\boldsymbol{j}_\pm$ is the ionic current for cations and anions, respectively.
The stress tensor comprises the hydrodynamic and Maxwell stress tensor as $\overleftrightarrow{\boldsymbol{\sigma}}=\overleftrightarrow{\boldsymbol{\sigma}_\mathrm{H}}+\overleftrightarrow{\boldsymbol{\sigma}_\mathrm{M}}$, which are defined by   
\begin{equation}
\overleftrightarrow{\boldsymbol{\sigma}_\mathrm{H}} = \left\{\begin{array}{ll}
\displaystyle -p\boldsymbol{I} + \eta_\mathrm{in}\left(\nabla\otimes \boldsymbol{u}+(\nabla\otimes \boldsymbol{u})^t\right) & \textrm{for } r<R\\\\
\displaystyle -p\boldsymbol{I} + \eta_\mathrm{out}\left(\nabla\otimes \boldsymbol{u}+(\nabla\otimes \boldsymbol{u})^t\right) & \textrm{for } r>R
\end{array}\right.,
\end{equation}
\begin{equation}
\overleftrightarrow{\boldsymbol{\sigma}_\mathrm{M}} = \left\{\begin{array}{ll}
\displaystyle -\frac{\varepsilon_\mathrm{in}\varepsilon_0}{2}(\nabla\psi)^2\boldsymbol{I} + \varepsilon_\mathrm{in}\varepsilon_0\left(\nabla\psi\otimes\nabla\psi\right) & \textrm{for } r<R\\\\
\displaystyle -\frac{\varepsilon_\mathrm{out}\varepsilon_0}{2}(\nabla\psi)^2\boldsymbol{I} + \varepsilon_\mathrm{out}\varepsilon_0\left(\nabla\psi\otimes\nabla\psi\right) & \textrm{for } r>R
\end{array}\right.,
\end{equation}
where $\boldsymbol{I}$ is the unit tensor. 
The charge density is defined by 
\begin{equation}
\rho(\boldsymbol{r}) =  \left\{\begin{array}{ll}
e\left(c_\mathrm{+,in}-c_\mathrm{-,in}\right) & \textrm{for } r<R \\
e\left(c_\mathrm{+,out}-c_\mathrm{-,out}\right) & \textrm{for } r>R
\end{array}\right.,
\end{equation}
where $e$ is the elementary charge and $c_\mathrm{\pm,in}$ and $c_\mathrm{\pm,out}$ are the local cation and anion concentrations inside or outside the droplet.
Although ionic currents consist of convection and conduction, a linear approximation of the surface charge density and the surface potential does not require explicit equations for ionic currents \cite{Henry1931,Booth1951,OBrien1978}. 

The boundary conditions for electrophoresis are 
\begin{eqnarray}
\boldsymbol{u} \cdot \boldsymbol{e}_r |_{r=R} & = & 0,\label{eq:9} \\
\boldsymbol{u} \cdot \boldsymbol{e}_\theta|_{r=R+0} -   \boldsymbol{u} \cdot \boldsymbol{e}_\theta|_{r=R-0} & = & 0, \label{eq:10}\\ 
\sigma_{r\theta}|_{r=R+0}-\sigma_{r\theta}|_{r=R-0} & = & 0, \label{eq:11} \\
\boldsymbol{j}_\pm\cdot \boldsymbol{e}_r|_{r=R} &= & 0 \label{eq:12} \\
\nabla p|_{r \to \infty} & = & 0,  \label{eq:13}\\
c_{i,\mathrm{out}}|_{r\to\infty} & = & c_\mathrm{out}^\infty, \label{eq:14}\\
\boldsymbol{u}|_{r \to \infty} & = & U\boldsymbol{e}_z, \label{eq:15}\\
\psi|_{r \to \infty} & = & -E r \cos\theta, \label{eq:16}
\end{eqnarray}
where $U$ is the external uniform velocity, $E$ is the externally applied electric field, and $c_\mathrm{out}^\infty$ is the salt concentration outside the droplet, which is substituted by zero later.  
Eq.~\ref{eq:9} is the fixed interface condition, eq.~\ref{eq:10} is the continuity of the tangential velocity at the interface, eq.~\ref{eq:11} is the continuity of the tangential stress at the interface, and eq.~\ref{eq:12} is the condition that the ions cannot penetrate the interface.
The boundary condition for $\sigma_{rr}$ is unnecessary because of its fixed interface \cite{Landau1987}. 
When the surface charge gradient is considered, an additional boundary condition is necessary for the balance between ion diffusion and migration in the tangential direction to the interface \cite{Saville1991,Hill2020,Hill2020_2,Hill2021}, which is further explained in Appendix.

When the electric field ($E= 0$) is switched off and the uniform flow $U$ is applied, the hydrodynamic equations and boundary conditions provide the velocity field and the frictional force of the liquid droplet in an external uniform flow, where the force is given by 
\begin{equation}
F_z = \int_{r=R} (\overleftrightarrow{\boldsymbol{\sigma}}\cdot d\boldsymbol{S})_z = \int_{r\to\infty} (\overleftrightarrow{\boldsymbol{\sigma}_\mathrm{H}}\cdot d\boldsymbol{S})_z = \gamma U,
\end{equation} 
where $\int_{r>R} \nabla\cdot\overleftrightarrow{\boldsymbol{\sigma}} d^3 r = 0$ and $\int_{r\to\infty}\overleftrightarrow{\boldsymbol{\sigma}}_\mathrm{M}\cdot d\boldsymbol{S}=0$ are used and 
\begin{equation}
\gamma = 2\pi\eta_\mathrm{out}R\frac{3\eta_\mathrm{in}+2\eta_\mathrm{out}}{\eta_\mathrm{in}+\eta_\mathrm{out}},
\end{equation}
is the Hadamard-Rybczynski friction coefficient \cite{Hadamard1911,Rybczynski1911}.
When the external uniform flow is switched off ($U = 0$) and the electric field $E$ is applied, the equations provide the force induced by the electric field. The effective charge $Q_\mathrm{eff}$ of the liquid droplet is then defined by the force divided by the electric field,
\begin{equation}
Q_\mathrm{eff} = \frac{F_z}{E} =\frac{1}{E} \int_{r\to\infty} (\overleftrightarrow{\boldsymbol{\sigma}}\cdot d\boldsymbol{S})_z. 
\end{equation}
If the surrounding liquid has no ions, $\int_{r\to\infty} (\overleftrightarrow{\boldsymbol{\sigma}_\mathrm{M}}\cdot d\boldsymbol{S})_z=QE$ does not disappear. However, when the surrounding liquid at least has counterions, $\int_{r\to\infty} (\overleftrightarrow{\boldsymbol{\sigma}_\mathrm{M}}\cdot d\boldsymbol{S})_z$ disappears. 
This difference is discussed later.
Consequently, because the electrophoresis is a linear problem, the electrophoretic mobility $\mu = -U/E$ is obtained by the force balance condition $F_z=0$ where $E$ and $U$ are the applied external fields.
Therefore, 
\begin{equation}
\mu = \frac{Q_\mathrm{eff}}{\gamma}.
\end{equation}
 
In the following sections, we calculate the equilibrium state and linear response within the approximation of a low surface charge and a low net charge (Debye-H\"uckel linearization). 

\subsection{Equilibrium state}
At equilibrium ($U=E=0$), all quantities are spherically symmetric, and the ions are distributed according to the Boltzmann distribution:
\begin{eqnarray}
c_{\pm,\mathrm{in}}(r) &=&  c^R_{\pm,\mathrm{in}} \mathrm{e}^{\mp(\Psi^\mathrm{eq}(r)-\Psi_R)},\\
c_{\pm,\mathrm{out}}(r) &=& c^\infty_{\mathrm{out}}\mathrm{e}^{\mp\Psi^\mathrm{eq}(r)},
\end{eqnarray}
where $c_\mathrm{\pm,\mathrm{in}}^R$ is the local concentration of cations or anions at the internal interface, $c^\infty_\mathrm{out}$ is the bulk ion concentration outside the droplet, $\Psi^\mathrm{eq}(r)=e\psi^\mathrm{eq}(r)/k_\mathrm{B}T$ is the dimensionless equilibrium electrostatic potential, and $\Psi_R = e\psi_R/k_\mathrm{B}T$ is the dimensionless potential at $r=R$. 
Note that $c_\mathrm{+,\mathrm{in}}^R$ and $c_\mathrm{-,\mathrm{in}}^R$ are generally different, and the partition equilibrium of the ion species across the interface is not considered. 
The equilibrium potential profile $\Psi^\mathrm{eq}(r)$ is calculated by the Poisson-Boltzmann equation
\begin{equation}
\begin{split}
&\frac{1}{r^2}\frac{d}{dr}\left[r^2\frac{d}{dr}\Psi^\mathrm{eq}(r)\right]=\\\\
&\left\{\begin{array}{ll}
-\kappa_{+,\mathrm{in}}^2\mathrm{e}^{-(\Psi^\mathrm{eq}-\Psi_R)}+\kappa_{-,\mathrm{in}}^2\mathrm{e}^{\Psi^\mathrm{eq}-\Psi_R} &\textrm{for } r<R\\
\kappa^2_\mathrm{out}\sinh\Psi^\mathrm{eq} &\textrm{for } r>R
\end{array}\right.,
\end{split}
\label{eq:23}
\end{equation}
where 
\begin{eqnarray}
\kappa_{i,\mathrm{in}}^2  & = & \frac{e^2 c_{i,\mathrm{in}}^R}{\varepsilon_\mathrm{in}\varepsilon_0 k_\mathrm{B}T},\\
\kappa^2_\mathrm{out} &=& \frac{2e^2 c^{\infty}_\mathrm{out}}{\varepsilon_\mathrm{out}\varepsilon_0 k_\mathrm{B}T}.
\end{eqnarray}
$\kappa_\mathrm{out}^{-1}$ is the thickness of the electric double layer outside the droplet.
The boundary condition is $d\Psi^\mathrm{eq}/dr|_{r=0}=0$, $\Psi^\mathrm{eq}|_{r\to\infty}=0$, $\Psi(R)=\Psi_R$.

The surface charge density is defined  by
\begin{equation}
\sigma_R = -\varepsilon_0\left[\left.\varepsilon_\mathrm{out}\frac{d\psi^\mathrm{eq}}{dr}\right|_{R+0}-\left.\varepsilon_\mathrm{in}\frac{d\psi^\mathrm{eq}}{dr}\right|_{R-0}\right],
\end{equation}
and the net charge of the droplet is defined by
\begin{equation}
\begin{split}
Q &= 4\pi R^2\left(-\varepsilon_\mathrm{in}\varepsilon_0 \left.\frac{d\psi^\mathrm{eq}}{dr}\right|_{r=R-0} +\sigma_R\right) \\
& =-4\pi R^2\varepsilon_\mathrm{out}\varepsilon_0 \left.\frac{d\psi^\mathrm{eq}}{dr}\right|_{r=R+0}
\end{split}
\end{equation}
where the surface charge density is assumed to be ion adsorption from the water phase rather than the oil phase.
The Poisson-Boltzmann equation \ref{eq:23} is linearized and solved as follows:
\begin{equation}
\begin{split}
&\Psi^\mathrm{eq}(r) = \\
&\left\{\begin{array}{ll}
\displaystyle \Psi_R +  \frac{\kappa_\mathrm{+,in}^2-\kappa_\mathrm{-,in}^2}{\kappa_\mathrm{in}^2}\left(1-\frac{R\sinh\kappa_\mathrm{in} r}{r\sinh\kappa_\mathrm{in}R}\right) & \textrm{for } r<R\\\\
\displaystyle \Psi_R \frac{R}{r} \mathrm{e}^{-\kappa_\mathrm{out}(r-R)} & \textrm{for } r>R
\end{array}\right.,
\end{split}
\label{eq:27}
\end{equation}
where $\kappa_\mathrm{in}^2=\kappa_{+,\mathrm{in}}^2+\kappa_{-,\mathrm{in}}^2$.
Within the limit of $c_\mathrm{out}^\infty\to 0$ (water-in-oil droplet), eq.~\ref{eq:27} for $r>R$ approaches $\Psi_R(R/r)$, which is the exact solution of the Poisson equation, and thus the condition of the small net charge ($|\Psi_R|\ll 1$) becomes unnecessary.  

\begin{figure*}
\includegraphics{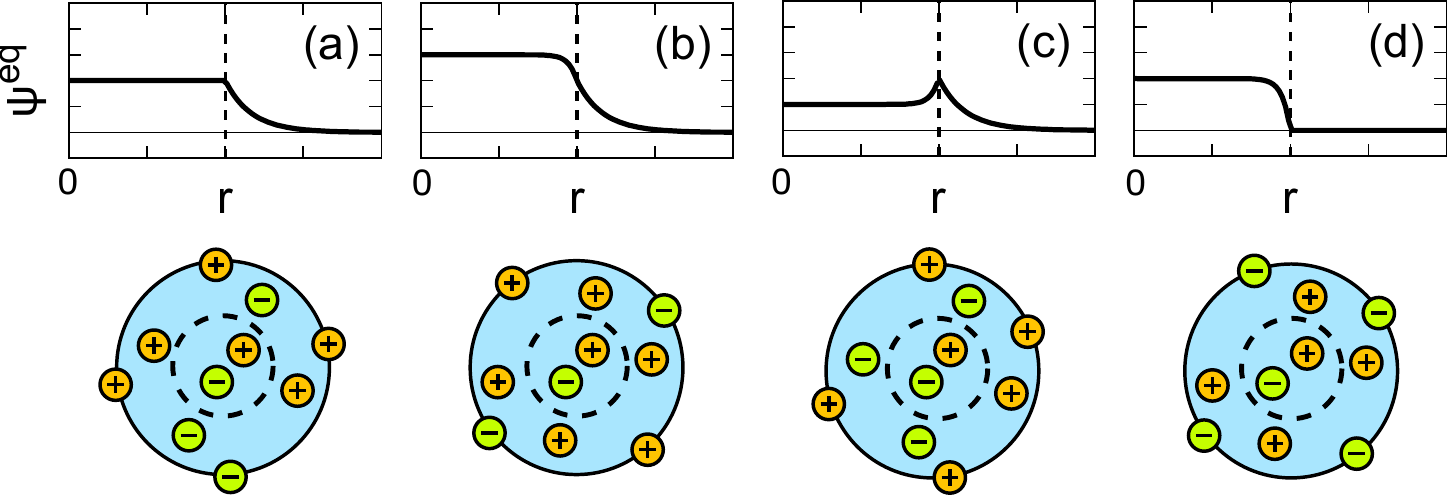}
\caption{
Typical profiles of $\psi^\mathrm{eq}(r)$ with $c_\mathrm{out}^\infty=0$ and $Q\ge 0$ and their schematic illustrations of the ion distribution.
The vertical broken lines are the location of the droplet interface and the horizontal lines are $\psi^\mathrm{eq}=0$.
(a) $\sigma_R=Q/4\pi R^2$,
(b) $\sigma_R<Q/4\pi R^2$,
(c) $\sigma_R>Q/4\pi R^2$.
(d) $Q=0$ and $\sigma_R<0$.
The ions on the droplet interface in the illustrations denote the surface charge, whereas the ions between the interface and inner dashed circles denote the electric double layer.
Inside the dashed circles, the ionic charge density is almost zero because the potential profile $\psi^\mathrm{eq}(r)$ is flat.
}
\label{fig:2}
\end{figure*}

Typical electrostatic profiles and their schematic illustrations are plotted in Fig.~\ref{fig:2}.
Because the surface charge density, $\sigma_R$, and the net charge, $Q$, are the independent parameters, qualitatively different profiles of the electrostatic potential exist for a water-in-oil droplet. 
When the charge of the droplet is distributed only at the interface (Fig.~\ref{fig:2}a), the potential is flat inside the droplet.
In (b), the rest of the droplet charge, $Q-4\pi R^2\sigma_R$, is positive, and the electric double layer forms inside the droplet, whereas in (c) the rest of the droplet charge is negative, and thus, the charge of the double layer inside is negative.
In (d), the net charge of the droplet is zero, and thus, the potential outside is flat. 
In the illustrations, the ions on the droplet interface denote the surface charge, whereas the ions between the interface and inner dashed circles denote the electric double layer.
Inside the dashed circles, the ionic charge density is almost zero because the potential profile $\psi^\mathrm{eq}(r)$ at this range is flat for all cases of (a) to (d).
Experimental classification of water-in-oil droplets or emulsions into these four cases shown in Fig.~\ref{fig:2} is still difficult because no reliable method to control and measure the surface charge density and net charge. 

\subsection{Linear response}
When an external uniform flow and/or an external electric field is applied, the deviations of all quantities from the equilibrium state become axisymmetric, and thus the velocity field can be described using the function $h(r)$ as \cite{OBrien1978} 
\begin{equation}
\boldsymbol{u}(r,\theta) = -\frac{2h(r)\cos\theta}{r}\boldsymbol{e}_r +\frac{\sin\theta}{r}\frac{d}{dr}[rh(r)]\boldsymbol{e}_\theta.
\label{eq:28}
\end{equation}
Note that the function $-rh(r)\sin^2\theta$ is the Stokes stream function.
The velocity field under an external uniform flow without an electric field was previously analytically solved \cite{Hadamard1911,Rybczynski1911}.
Therefore, in this work, the velocity field under an external electric field without an external uniform flow ($U=0$) is studied. 
As we are focusing on low surface charge densities and low net charges, it is unnecessary to explicitly consider the deviations of the ionic concentration fields from the equilibrium distribution \cite{Henry1931}; thus, eqs.~\ref{eq:4}, \ref{eq:5}, and \ref{eq:12} are not solved. 
Instead, these approximations impose the use of the electric potential profile given by 
\begin{equation}
\delta\psi(r,\theta) = \left\{\begin{array}{ll}
\displaystyle -(1+\lambda)Er\cos\theta & \textrm{for } r<R\\\\
\displaystyle -\left(1+\lambda\frac{R^3}{r^3}\right)Er\cos\theta  & \textrm{for } r>R
\end{array}\right..
\label{eq:29}
\end{equation} 
The coefficient $\lambda$ was classically understood as the difference in conductivity $(\sigma_\mathrm{out}-\sigma_\mathrm{in})/(2\sigma_\mathrm{out}+\sigma_\mathrm{in})$ where $\sigma_\mathrm{in}$ and $\sigma_\mathrm{out}$ are the conductivities of the inner and outer liquids \cite{Henry1931,Booth1951}. 
However, if the ions cannot penetrate the other phase (eq.~\ref{eq:12}), this coefficient is $\lambda=1/2$ and is independent of the conductivities $\sigma_\mathrm{in}$ and $\sigma_\mathrm{out}$, and the dielectric constants $\varepsilon_\mathrm{in}$ and $\varepsilon_\mathrm{out}$ \cite{Levine1973}.
For mercury droplets, the coefficients are different inside ($\lambda=-1$) and outside ($\lambda = 1/2$) due to the perfect polarization of the droplet \cite{Levine1973,Ohshima1984}. 
Using eqs.~\ref{eq:28} and \ref{eq:29}, the hydrodynamic equations, eqs.~\ref{eq:2} and \ref{eq:3}, are reduced to the equation for $h(r)$ as follows \cite{OBrien1978}:
\begin{equation}
L^2 h(r) = -f(r)E,
\label{eq:30}
\end{equation}
where
\begin{equation}
L = \frac{d}{dr}\left(\frac{1}{r^2}\frac{d}{dr}r^2\right) = \frac{d^2}{dr^2}+\frac{2}{r}\frac{d}{dr}-\frac{2}{r^2}, 
\end{equation}
and 
\begin{equation}  
\begin{split}
f(r) &= \left\{\begin{array}{ll}
\displaystyle \frac{\varepsilon_\mathrm{in}\varepsilon_0 \kappa_\mathrm{in}^2 }{\eta} (1+\lambda) \frac{d\psi^\mathrm{eq}}{dr} & \textrm{for }r<R\\\\
\displaystyle \frac{\varepsilon_\mathrm{out}\varepsilon_0 \kappa_\mathrm{out}^2 }{\eta} \left(1+\lambda\frac{R^3}{r^3}\right)\frac{d\psi^\mathrm{eq}}{dr} & \textrm{for }r>R
\end{array}\right..
\end{split}
\end{equation}
The solution of eq.~\ref{eq:30} is
\begin{equation}
\begin{split}
&h(r)=\\
&\left\{\begin{array}{l}
\displaystyle A_1 \frac{r}{R} + A_2\frac{R^2}{r^2} + A_3 \frac{r^3}{R^3}+A_4 - \int^r_0 G(r,r') f(r')Edr' \\ \textrm{for } r<R\\\\
\displaystyle B_1 \frac{r}{R} + B_2\frac{R^2}{r^2} + B_3 \frac{r^3}{R^3}+B_4 - \int^r_\infty G(r,r') f(r')Edr'\\ \textrm{for } r>R\\
\end{array}\right.,
\end{split}
\end{equation}
where $A_n$ and $B_n$ are constants and 
\begin{equation}
G(r,r')=-\frac{r'^5}{30r^2}+\frac{r'^3}{6}-\frac{rr'^2}{6}+\frac{r^3}{30}.
\end{equation}
The boundary conditions, eqs.~\ref{eq:9}--\ref{eq:11} and \ref{eq:15}--\ref{eq:16} for $(U,E)=(0,E)$, can be converted to $B_1=B_3=0$, $A_2=A_4=0$, $B_2+B_4-D_0=0$, $A_1+A_3-C_0=0$, $A_1+3A_3-C_1 = -2B_2-D_1$, and 
\begin{equation}
6\eta_\mathrm{in}A_3-\eta_\mathrm{in}C_2 = 6\eta_\mathrm{out}B_2-\eta_\mathrm{out}D_2 -\sigma_R(1+\lambda)R^2,
\end{equation}
where $C_n$ and $D_n$ are defined as
\begin{eqnarray}
\displaystyle C_n &=& R^n\int^R_0 \left(\frac{d}{dR}\right)^n G(R,r) f(r)dr, \\
\displaystyle D_n &=& R^n\int^R_\infty \left(\frac{d}{dR}\right)^n G(R,r) f(r)dr.
\end{eqnarray}
These equations yield $B_4$ as
\begin{equation}
\begin{split}
B_4 & = \frac{\eta_\mathrm{in}(3C_0-3C_1+C_2+6D_0+3D_1)}{6(\eta_\mathrm{in}+\eta_\mathrm{out})} \\
& + \frac{\eta_\mathrm{out}(6D_0-D_2)-\sigma_R(1+\lambda)R^2}{6(\eta_\mathrm{in}+\eta_\mathrm{out})}\label{eq:37}.
\end{split}
\end{equation}
Because the hydrodynamic force is obtained as $8\pi\eta_\mathrm{out}B_4$ \cite{OBrien1978}, the effective charge $Q_\mathrm{eff}$ is 
\begin{equation}
Q_\mathrm{eff} = Q + 8\pi\eta_\mathrm{out}B_4, 
\end{equation}
where $Q$ is due to long-range Maxwell stress when the surrounding liquid contains no ions (even counterions).
In this case, $\kappa_\mathrm{out}=0$, and thus, $f(r)=0$ for $r>R$.   
Conversely, if the surrounding liquid contains counterions (this corresponds to $\kappa_\mathrm{out}\to 0$ not $\kappa_\mathrm{out}=0$)), $Q=0$ is obtained because of the Gauss's law in electrostatics. 
Later, we demonstrate that these two limiting situations provide the same electrophoretic mobility.

\begin{figure}
\begin{center}
\includegraphics{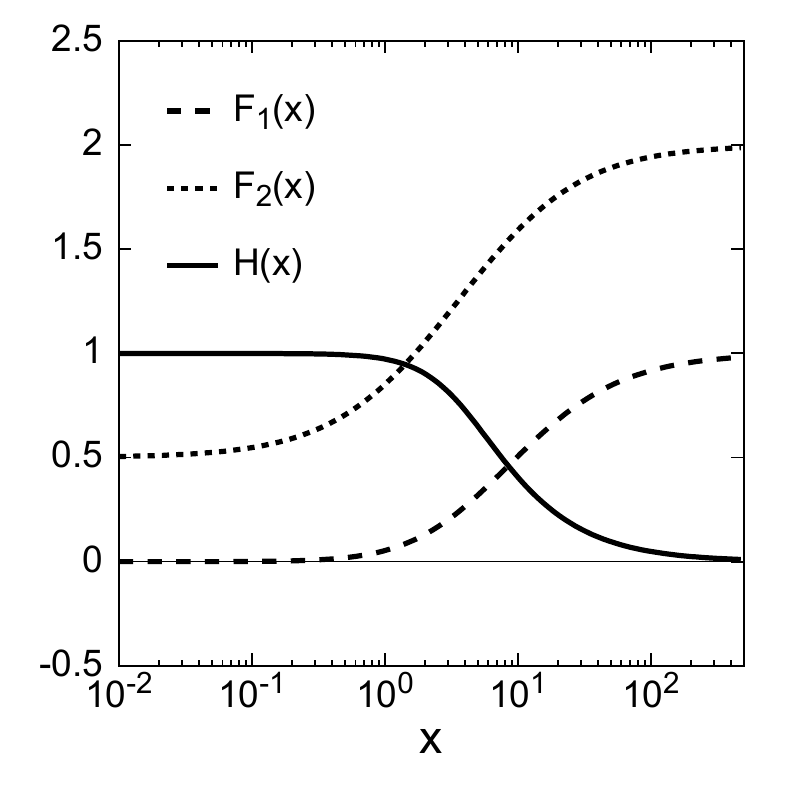}
\caption{
Shape of functions $F_1(x)$, $F_2(x)$, and $H(x)$. 
}
\label{fig:1}
\end{center}
\end{figure}

The resulting electrophoretic mobility $\mu=Q_\mathrm{eff}/\gamma$ is broken down into five parts:
\begin{equation}
\mu = \mu_1 + \mu_2 + \mu_3 + \mu_4 + \mu_5, 
\end{equation}
where $\mu_1$ to $\mu_4$ correspond to $3C_0-3C_1+C_2$, $6D_0+3D_1$, $6D_0-D_2$, and $-\sigma_R(1+\lambda)R^2$ terms in eq.~\ref{eq:37}, and $\mu_5$ corresponds to $Q/\gamma$.
Each term is explicitly obtained as 
\begin{eqnarray}
\mu_1 & = & -\frac{\varepsilon_\mathrm{out}\varepsilon_0(1-H(\kappa_\mathrm{in}R))}{3\eta_\mathrm{in}+2\eta_\mathrm{out}}\left[(1+\kappa_\mathrm{out}R)\psi_R-\frac{R\sigma_R}{\varepsilon_\mathrm{out}\varepsilon_R}\right],\nonumber\\\\
\mu_2 & = & \frac{2\eta_\mathrm{in}\varepsilon_\mathrm{out}\varepsilon_0\psi^\mathrm{eq}_R}{\eta_\mathrm{out}(3\eta_\mathrm{in}+2\eta_\mathrm{out})}\left[1+ \frac{F_1(\kappa_\mathrm{out}R)}{2}\right], \label{eq:41}\\
\mu_3 & = &  \frac{\varepsilon_\mathrm{out}\varepsilon_0\psi^\mathrm{eq}_R}{3\eta_\mathrm{in}+2\eta_\mathrm{out}} \left[(1+\kappa_\mathrm{out} R)+\frac{4}{3}-\frac{2F_2(\kappa_\mathrm{out}R)}{3}\right], \label{eq:42}\\
\mu_4 & = & -\frac{R\sigma_R}{3\eta_\mathrm{in}+2\eta_\mathrm{out}},\label{eq:43}\\
\mu_5 & = & \frac{2\varepsilon_\mathrm{out}\varepsilon_0 \psi_R}{ 3\eta_\mathrm{in}+2\eta_\mathrm{out} } \left[ 1+\frac{\eta_\mathrm{in}}{\eta_\mathrm{out}}\right].\label{eq:44}
\end{eqnarray}
We define the functions according to the work of Booth \cite{Booth1951} as  
\begin{eqnarray}
F_1(x) & = & 1+6E_5(x)\mathrm{e}^x-15E_7(x)\mathrm{e}^x, \\
F_2(x) & = & 2-6E_5(x)\mathrm{e}^x, \\
F_3(x) & = & (\mathrm{e}^x(x-1)+\mathrm{e}^{-x}(x+1))^{-1}, \\    
F_4(x) & = & \mathrm{e}^x(x^2-3x+3)-\mathrm{e}^{-x}(x^2+3x+3), 
\end{eqnarray}
where $E_n(x) = \int^\infty_1\mathrm{e}^{-xt}t^{-n}dt$.
Additionally, a function is newly introduced by
\begin{equation}
H(x) =  5F_3(x)F_4(x)/x^2.
\end{equation}
The shapes of the functions $F_1(x)$, $F_2(x)$, and $H(x)$ are plotted in Fig.~\ref{fig:1}.

Each electrophoretic mobility term is described below: 
$\mu_1$ is the contribution from the double layer inside the droplet because this originates from $C_n$. 
$\mu_2$ is the contribution from the double layer outside the droplet, and in the solid particle limit ($\eta_\mathrm{in}/\eta_\mathrm{out}\to\infty$), only this term survives and approaches Henry's formula \cite{Henry1931}.
$\mu_3$ is another contribution from the double layer outside the droplet, but it only exists for the liquid droplet. 
For mercury droplets, this contribution is dominant \cite{Levich1962,Levine1973,Ohshima1984}. 
$\mu_4$ is the contribution from the continuity of the Maxwell tensor, which vanishes for mercury droplets  \cite{Levich1962,Levine1973,Ohshima1984}. 
In Booth's work \cite{Booth1951}, the effect of surface charge density on the electrophoretic mobility was mentioned, but no explicit mobility equation containing $\sigma_R$ was derived.
In addition, at that time, the coefficient $\lambda$ was considered $(\sigma_\mathrm{out}-\sigma_\mathrm{in})/(2\sigma_\mathrm{out}+\sigma_\mathrm{in})$ \cite{Henry1931, Booth1951}, but $\lambda=1/2$ is used here. 

The last term $\mu_5$ is the contribution from the net charge of the droplet, which is unnecessary if the surrounding liquid has counterions. 
From eqs.~\ref{eq:41}, \ref{eq:42}, and \ref{eq:44}, we obtain 
\begin{equation}
(\mu_2+\mu_3)|_{\kappa_\mathrm{out}R\to 0} = \mu_5,
\end{equation} 
indicating that both cases, $\kappa_\mathrm{out}\to 0$ and $\kappa_\mathrm{out}=0$, have the same electrophoretic mobility. 
Therefore, it is unnecessary to consider the two situations, which are hereafter referred to as water-in-oil droplet, separately.
The surface potential $\psi_R$ of a water-in-oil droplet can be calculated by using the net charge $Q$ as
\begin{equation}
\psi_R = \frac{Q}{4\pi\varepsilon_\mathrm{out}\varepsilon_0 R}.
\end{equation} 
Therefore, we can rewrite the electrophoretic mobility of the water-in-oil droplet, $\mu=\mu_1+\mu_4+\mu_5$, using $\sigma_R$ and $Q$ as
\begin{equation}
\mu = \frac{Q/2\pi R}{3\eta_\mathrm{in}+2\eta_\mathrm{out}}\left[\frac{1}{2}+\frac{\eta_\mathrm{in}}{\eta_\mathrm{out}}+\frac{H(\kappa_\mathrm{in}R)}{2}\right]-\frac{\sigma_R R H(\kappa_\mathrm{in}R)}{3\eta_\mathrm{in}+2\eta_\mathrm{out}}. \label{eq:52}
\end{equation}
If we multiply $\mu$ by $\gamma$, we obtain the effective charge 
\begin{equation}
Q_\mathrm{eff} = Q\frac{2\eta_\mathrm{in}+\eta_\mathrm{out}(1+2H(\kappa_\mathrm{in}R))}{2(\eta_\mathrm{in}+\eta_\mathrm{out})} - 2\pi R^2\sigma_R \frac{\eta_\mathrm{out} H(\kappa_\mathrm{in}R)}{\eta_\mathrm{in}+\eta_\mathrm{out}}. \label{eq:53} 
\end{equation}
Eqs.~\ref{eq:52} and \ref{eq:53} clearly demonstrate that the electrophoretic mobility of a water-in-oil droplet is separately affected by the net charge $Q$ and the surface charge density $\sigma_R$.  
In the next section, we discuss the results using eqs.~\ref{eq:52} and \ref{eq:53}.

Before proceeding to the following section, we discuss the validity of assuming that the oil is ion-free \cite{Labib1988,Kosmulski1999,Fuoco1992,Hung1980, Onuki2011}.
The ion partition between the water and oil phases depends on the chemical properties of the ions and the oil \cite{Fuoco1992,Hung1980,Onuki2011}.
For example, the Gibbs transfer energy of alkali metals and halide ions from water to 1,2-dichloroethane (dielectric constant $\sim 10$) is approximately 30--50 kJ/mol (12--20$k_\mathrm{B}T$) \cite{Fuoco1992}. This indicates that the ion concentration in the oil phase is significantly lower than that in the aqueous phase where the Boltzmann factor is at most $10^{-5}$. 
In addition to these ions, it is assumed that some of the ionic surfactants are not dissolved in the low-dielectric oil phase.
For example, the Gibbs transfer energy of dodecylsulfate ions from water to nitrobenzene (dielectric constant $\sim 35$) is $+5\,$kJ/mol ($2k_\mathrm{B}T$) \cite{Hung1980}, and the HLB of sodium dodecylsulfate is estimated by $\sim 40$ \cite{Davies1957}.
Due to the different Gibbs transfer energies of the cations and anions in the water droplet, a Galvani potential difference is created \cite{Onuki2011}, which can contribute to electrophoretic mobility \cite{Uematsu2015}.
However, in this study, we do not consider this contribution for simplicity.
Ion penetration at the interface is unlikely because the conductivity of the oil phase is negligible.

\section{Results and discussion}
\subsection{Limiting equations for mobility}
First, we take the limit of $\kappa_\mathrm{in}R\to 0$ (small droplet) and $\kappa_\mathrm{in}R\to\infty$ (large droplet) of eq.~\ref{eq:53} using the relations, $H(x\to 0)=1-x^2/35$ and $H(x\to\infty)=5/x$. Then, we obtain
\begin{eqnarray}
Q_\mathrm{eff}|_{\kappa_\mathrm{in}R\to 0} & = & Q\frac{2\eta_\mathrm{in}+3\eta_\mathrm{out}}{2(\eta_\mathrm{in}+\eta_\mathrm{out})}-2\pi R^2\sigma_R\frac{\eta_\mathrm{out}}{\eta_\mathrm{in}+\eta_\mathrm{out}},\label{eq:n56}\\ 
Q_\mathrm{eff}|_{\kappa_\mathrm{in}R\to \infty} & = & Q\frac{2\eta_\mathrm{in}+\eta_\mathrm{out}}{2(\eta_\mathrm{in}+\eta_\mathrm{out})}-\frac{10\pi R\sigma_R}{\kappa_\mathrm{in}}\frac{\eta_\mathrm{out}}{\eta_\mathrm{in}+\eta_\mathrm{out}}.\nonumber\\
 \label{eq:55} 
\end{eqnarray}
When $\sigma_R = Q/4\pi R^2$, eq.~\ref{eq:n56} approaches to $Q_\mathrm{eff}=Q$ like H\"uckel formula whereas the second term in eq.~\ref{eq:55} vanishes.
Therefore, eq.~\ref{eq:55} suggests that the effective charge of a large water droplet in low-molecular-weight insulating oil ($\eta_\mathrm{out}\approx\eta_\mathrm{in}$) is $Q_\mathrm{eff}=(3/4)Q$.

The viscosity of the oil can be changed by choosing the molecular weight of the oil. 
For example, at room temperature, benzene and toluene are less viscous than water $\sim (2/3)\eta_\mathrm{water}$ where $\eta_\mathrm{water}$ is the water viscosity, whereas the viscosity of silicone oil ranges from $\eta_\mathrm{water}/2$ to $10^6\eta_\mathrm{water}$.
In addition, our theory can be applied to water droplets (even oil droplets) in the gas phase \cite{Millikan1911}. 
At room temperature, air has a viscosity of about $\eta_\mathrm{water}/50$. 
Thus, the two limits of  $\eta_\mathrm{out}/\eta_\mathrm{in}\to 0$ (droplet in gas \cite{Millikan1911}) and $\eta_\mathrm{out}/\eta_\mathrm{in}\to \infty$ (viscous oil) are worth considering. Thus, we obtain
\begin{eqnarray}
 Q_\mathrm{eff}|_{\eta_\mathrm{out}/\eta_\mathrm{in}\to 0}  & = & Q, \\
 Q_\mathrm{eff}|_{\eta_\mathrm{out}/\eta_\mathrm{in}\to \infty} & = & Q\frac{1+2H(\kappa_\mathrm{in}R)}{2}-2\pi R^2\sigma_R H(\kappa_\mathrm{in}R).\nonumber\\ 
\end{eqnarray}
Another important limit is large or small droplets surrounded by viscous oil, and in this limit 
\begin{eqnarray}
Q_\mathrm{eff}|_{\eta_\mathrm{out}/\eta_\mathrm{in}\to \infty, \kappa_\mathrm{in}R\to 0 } & = & \frac{3Q}{2}-2\pi R^2\sigma_R,\label{eq:n60}\\
Q_\mathrm{eff}|_{\eta_\mathrm{out}/\eta_\mathrm{in}\to \infty, \kappa_\mathrm{in}R\to \infty} & = &\frac{Q}{2}-\frac{10\pi R\sigma_R}{\kappa_\mathrm{in}}. \label{eq:56}
\end{eqnarray}
When $\sigma_R = Q/4\pi R^2$, eq.~\ref{eq:n60} approaches to $Q_\mathrm{eff}=Q$ like H\"uckel formula whereas the second term in eq.~\ref{eq:56} vanishes.
Therefore, eq.~\ref{eq:56} suggests that the effective charge of a large water droplet in high-viscous insulating oil is a half of the net charge, $Q_\mathrm{eff}=Q/2$.

Subsequently, we consider the case of $Q=0$ as shown in Fig.~\ref{fig:2}d.
In principle, the net charge can leak into the external liquid by the relaxation time $\tau = \varepsilon_\mathrm{out}\varepsilon_0/\sigma_\mathrm{out}$.
Substituting typical values for insulating oils, $\varepsilon_\mathrm{out}=2$ and $\sigma_\mathrm{out} = 10^{-13}\,$S/m, we obtain $\tau\sim 100\,$s \cite{Esmaeil2012}.
Therefore, within minutes of preparing the emulsions and droplets, the net charge will be negligible.
The electrophoretic mobility in this case is
\begin{equation}
\mu = -\frac{\sigma_R RH(\kappa_\mathrm{in}R)}{3\eta_\mathrm{in}+2\eta_\mathrm{out}},\label{eq:60}
\end{equation} 
indicating that the electrophoretic mobility has the inverse sign of the surface charge density. 
With the limit $\kappa_\mathrm{in}R\to\infty$, eq.~\ref{eq:60} reduces to 
\begin{equation}
\mu|_{\kappa_\mathrm{in}R\to\infty} = -\frac{5\sigma_R}{\kappa_\mathrm{in}(3\eta_\mathrm{in}+2\eta_\mathrm{out})}\label{eq:61}
\end{equation}
Because in this limit $\sigma_R = -\varepsilon_\mathrm{in}\varepsilon_0 \kappa_\mathrm{in}\psi_0$ where $\psi_0$ is the potential at the center, eq.~\ref{eq:61} reduces 
\begin{equation}
\mu= \frac{5\varepsilon_\mathrm{in}\varepsilon_0\psi_0}{3\eta_\mathrm{in}+2\eta_\mathrm{out}},
\end{equation}
indicating that the double layer inside the droplet adversely affects electrophoresis, which has not been pointed out thus far to the best of our knowledge. 

\subsection{Discussion of the experimental data in the literature}

First, we test our theory using the experimental data obtained by Albers and Overbeek \cite{Overbeek1959-1}.
They prepared a water-in-benzene emulsion using a metallic soap as a stabilizer, and the radius was about $1\,\mu$m to $4\,\mu$m. 
Metallic soaps generally dissolve better in organic liquids than in water. 
However, as metallic soaps do not dissociate in the oil phase more than in the aqueous phase, the dissociated multivalent cations and fatty acids may be better distributed in the water phase than the oil phase.
Therefore, we assume that our theory can be applied to this experiment.
 
For simplicity, we assume $\eta_\mathrm{in}=\eta_\mathrm{out}=1\,$mPa$\cdot$s, $R = 3\,\mu$m, $\varepsilon_\mathrm{in}=80$, $\varepsilon_\mathrm{out}=2$.
Water-in-benzene emulsions stabilized by various metallic oleates exhibit positive electrophoretic mobilities about $+10^{-10}\,$m$^2$/V$\cdot$s ($\zeta = +10\,$mV based on the H\"uckel formula $\mu = 2\varepsilon_\mathrm{out}\varepsilon_0\zeta/3\eta_\mathrm{out}$) and the effective charge is $Q_\mathrm{eff}=5\times 10^{-18}\,$C $= 30e$.
Because $R=3\,\mu$m and the maximum Debye length of the $\mathrm{CO_2}$-saturated water is $300\,$nm, we set $H(\kappa_\mathrm{in}R)=1/2$ according to Fig.~\ref{fig:1}. Then, the surface charge density is obtained as $\sigma_R = -0.3\,\mu$C/m$^2$ from eq.~\ref{eq:60}, which is quantitatively realistic.  
Here, we assume $Q=0$ because the net charge of the emulsion droplet is seemed to be relaxed by the leak and partition of charges.
Therefore, the positive zeta potential observed for a water-in-benzene droplet stabilized by metallic soaps can be explained by the adsorption of soap anions at the interface.
Because our theory neglects the Marangoni effect, this explanation is not conclusive, and other explanations are also possible, for example, strong partition of multivalent cations in the water droplet, or incomplete dissociation of metallic soaps with multivalent cations \cite{Overbeek1959-1}. 
We propose an experiment to validate our theory using a water-in-oil droplet stabilized by small amount of monovalent anionic and cationic surfactants that dissolve or dissociate only in the aqueous phase (HLB is high).
 
Second, we examine recent experiments on the electrophoretic mobility of surfactant-free water-in-oil droplets prepared by pipette injection \cite{Kang2008,Im2011,Mesquida2013,Choi2013,Im2020}. 
Before the droplet comes into contact with the electrode, its electrophoretic mobility is usually positive \cite{Mesquida2013,Choi2013}.
The origin of this positive electrophoretic mobility is believed to be the net charge that compensates for the negative surface charge on the pipette tip surface \cite{Choi2013,Mishra2020}.
In the work by Schoeler et.~al \cite{Mesquida2013}, $Q_\mathrm{eff}\sim 1\,$pC for a water droplet in silicone and paraffin oil ($\eta_\mathrm{out}/\eta_\mathrm{in}\sim 100$) with radius $R = 0.6\,$mm, whereas Choi et.~al observed the effective charge $Q_\mathrm{eff}\sim 100\,$pC for a water droplet in silicone oil ($\eta_\mathrm{out}/\eta_\mathrm{in}\sim 50$) with radius $R\approx 1\,$mm.
After the droplet comes into contact with the electrode, the water droplet becomes charged with the sign of the electrode. 
When a fully conductive sphere comes into contact with the electrode, the charge acquired by the sphere is $Q_\mathrm{M} = (\pi^2/6)(4\pi R^2)\varepsilon_\mathrm{out}\varepsilon_0 E$ \cite{Perez2002}.
Thus far, Choi et.~al measured the net charge of pipette-injected water droplets by the Faraday cup method, and they obtained $Q_\mathrm{eff} = Q$ for deionized water droplets using the Hadamard-Rybczynski friction coefficient \cite{Choi2013}.
Furthermore, Im investigated the wall effect on hydrodynamic friction and obtained $Q_\mathrm{eff}=Q_\mathrm{M}$ using the wall-effect-included Stokes friction \cite{Im2020}.
Our theory, eq.~\ref{eq:56}, predicts $Q_\mathrm{eff}=Q/2$ when the surface charge of a droplet is negligible, and thus the experimental data disagree with our theory.  
However, for a water droplet, the charge extracted from the electric current signal is usually smaller than $Q_\mathrm{M}$ \cite{Im2011, Ristenpart2017}.
The hydrodynamic friction usually deviates from the Hadamard-Rybczynski friction coefficient because of the wall effect \cite{Mesquida2013, Im2020} and unexpected impurities at the interface \cite{Hamlin2012,Im2020}.  
Impurities at the oil-water and air-water interfaces have been proposed by other studies to explain various experimental phenomena \cite{Uematsu2020,UematsuCOE,Roger2012}.
If our theory is applied to the Stokes friction (no slip boundary), the results are completely replaced by Henry's classical result \cite{Henry1931}. 
Eqs. \ref{eq:10} and \ref{eq:11} are replaced by $\boldsymbol{u} \cdot \boldsymbol{e}_\theta|_{r=R+0} = 0$ and $\boldsymbol{u} \cdot \boldsymbol{e}_\theta|_{r=R-0} = 0$; therefore, the inner flow of the droplet is independent of the outer flow. 
If the surrounding liquid contains no ions, even if the radius of the droplet is as large as $R\approx 1\,$mm, $Q_\mathrm{eff}=Q$ recovers.
In the context of the electrophoresis of water-in-oil droplet, $Q_\mathrm{eff}=Q$ was an assumption, and it was not examined itself \cite{Im2011,Hamlin2012,Choi2013,Mesquida2013}.
Therefore, to experimentally validate $Q_\mathrm{eff}=Q/2$, the experimental setup should be designed to prevent contamination and to provide independent measurements of the hydrodynamic friction and the electrostatic net charge of a millimeter-scale droplet.

\section{Conclusions}
We analytically calculated the electrophoretic mobility of a water-in-oil droplet with Debye-H\"uckel approximation. 
The resulting electrophoretic mobility depends separately on the net charge of the droplet and the surface charge density at the droplet interface. 
When the net charge is negligible, the electrophoretic mobility is proportional to the surface charge density with a negative coefficient. This indicates that the internal electric double layer adversely contributes to electrophoresis. 

Our theory is used to discuss the experimental data of water-in-oil emulsions and droplets in the literature.
The positive electrophoretic mobility of a water-in-benzene emulsion stabilized by metallic soaps can be explained by the negative surface charge of the absorbed soap ions. 
Recent electrophoretic mobility measurements of water-in-oil droplets suggest that the effective charge deduced from electrophoresis is identical to the net charge of the droplet. Conversely, our theory indicates that the effective charge should be half the net charge when the surface charge density is negligible.
However, in experiments, the Hadamard-Rybczynski friction is quite difficult to achieve, and the wall effect and unexpected impurities at the interface often modify the hydrodynamic friction \cite{Mesquida2013,Im2020}. 
The accurate electric measurement of the net charge of the droplet is another issue when comparing the effective electrophoretic charge with the net charge \cite{Ristenpart2017}.
Thus, it is still unclear what determines the electrophoretic mobility of water-in-oil droplets in experiments.

Finally, we briefly mention the outlook on the electrophoresis of water-in-oil droplets. 
In this paper, ion adsorption was not included in the model. 
However, surface charge density is generally caused by ion adsorption \cite{Uematsu2021}.
Ion partition to the oil phase would also be important because it creates a Galvani potential difference across the interface, even when the oil phase concentration is extremely low \cite{Onuki2011,Uematsu2015}. 
To include such effects in the model, the Poisson-Boltzmann and hydrodynamic equations must be treated completely nonlinearly.  
Marangoni flow, the stagnant cap of the surfactants at the interface, and droplet deformation are other important effects on electrophoresis \cite{Hamlin2012,Hill2020,Taylor1966}.

\vspace{5mm}
\section*{acknowledgement}
This work was supported by JST, PRESTO Grant Number JPMJPR21O2, Japan.
YU thanks JSPS KAKENHI Grant Numbers 18KK0151 and 20K14430, Sasakawa Scientific Research Grant from The Japan Science Society (2021-3001), Grant from  Kurita Water and Environment Foundation (21E006), and Qdai-jump Research Program from Kyushu University (R3-01302) for their support. 

\section{Appendix}
In this appendix, we briefly discuss how the Marangoni effect modifies the steady-state low-Reynolds-number electrohydrodynamic equations \cite{Saville1991}. 
The ionic currents inside the droplet and at the interface are given by
\begin{eqnarray}
\boldsymbol{j}_\pm & = & c_{\pm,\mathrm{in}} \boldsymbol{u}-\omega_\pm c_{\pm,\mathrm{in}} \nabla \mu_\pm, \\
\boldsymbol{j}_\pm^\mathrm{s} & = & \Gamma_\pm\boldsymbol{u}^\mathrm{s}-\omega_\pm^\mathrm{s}\Gamma_\pm\nabla^\mathrm{s}\mu^\mathrm{s}_\pm
\end{eqnarray}
where $\omega_\pm$ and $\omega_\pm^\mathrm{s}$ are the ionic mobilities inside the droplet and at the interface, and $\Gamma_\pm$ is the surface concentration of each ions.
The chemical potentials $\mu_\pm$ and $\mu_\pm^\mathrm{s}$ are defined by
\begin{eqnarray}
\mu_\pm & = & k_\mathrm{B}T \ln c_{\pm,\mathrm{in}} \pm e\psi,\\
\mu^\mathrm{s}_\pm & = & k_\mathrm{B}T\ln(\Gamma_\pm/z^*)\pm e \psi + \Delta G_\pm,
\end{eqnarray}
where $\Delta G_\pm$ is the adsorption energy, and $z^*$ is the thickness of the adsorption layer.
The equilibrium adsorption isotherm of ions can be obtained by the relation $\mu_\pm|_{r=R-0} = \mu_\pm^\mathrm{s}$, and this yields
\begin{equation}
\Gamma_\pm = K_\pm c_\pm|_{r=R-0},
\end{equation}
where $K_\pm = z^*\mathrm{e}^{-\Delta G_\pm / k_\mathrm{B}T}$ is the adsorption coefficient.
The adsorption-desorption dynamics can be included by the reaction equations \cite{Levich1962}, while in this paper we assume it is rapid according to the previous work \cite{Saville1991}.
Then, the surface charge density is given by 
\begin{equation}
\begin{split}
\sigma_R & = e(\Gamma_+-\Gamma_-)\\
& = e(K_+c_+|_{r=R-0}-K_-c_-|_{r=R-0}),
\end{split}
\end{equation}
and thus, under the assumption of rapid adsorption-desorption dynamics, the surface charge gradient is directly coupled to the distortion of the electric double layer.  
The conservation law for the ions at the interface is, therefore,
\begin{equation}
\nabla^\mathrm{s}\cdot \boldsymbol{j}^\mathrm{s}_\pm+(\boldsymbol{j}_\pm|_{r=R+0} - \boldsymbol{j}_\pm|_{r=R-0}) \cdot \boldsymbol{e}_r=0
\end{equation}
and for water-in-oil droplet (no ion outside the droplet) with rapid adsorption-desorption equilibrium we have
\begin{eqnarray}
\boldsymbol{j}_\pm\cdot \boldsymbol{e}_r |_{r=R+0}&=& 0, \\
\boldsymbol{j}_\pm \cdot  \boldsymbol{e}_r |_{r=R-0}& = & \nabla^\mathrm{s} \cdot \boldsymbol{j}^\mathrm{s}_\pm, 
\label{eq:73}
\end{eqnarray}
instead of eq.~\ref{eq:12}. 
The boundary condition of the tangential stress tensor is 
\begin{equation}
\sigma_{r\theta}|_{r=R+0}-\sigma_{r\theta}|_{r=R-0}  = k_\mathrm{B}T\sum_{i=\pm}(\nabla^\mathrm{s} \Gamma_i)\cdot \boldsymbol{e}_\theta 
\label{eq:74}
\end{equation}
instead of eq.~\ref{eq:11}.
To verify the assumption of neglecting the Marangoni effect (right hand side of eqs. \ref{eq:73} and \ref{eq:74}), it is necessary to solve full equations with nonlinrar electrostatics and to evaluate each term in specific parameter sets \cite{Saville1991}.
Because the Marangoni effect is usually proportional to the Marangni number $\Gamma_\pm R/\eta_\mathrm{in}\omega^\mathrm{s}$, it can be neglected when the surface concentration is small as $\Gamma_\pm \ll \eta_\mathrm{in}\omega^\mathrm{s}/R$.
For oil-in-water droplets with anionic surfactants, full calclations and verifications were performed for MHz frequency \cite{Hill2020,Hill2020_2}. 
However, for the water-in-oil droplets, a clear analytical or numerical study to explain how the Marangoni effect modifies the electrophoretic mobility is still lack, and thus, it remains a future problem.

\bibliography{droplet}

\end{document}